\documentclass[conference]{IEEEtran}
\IEEEoverridecommandlockouts
\usepackage{cite}
\usepackage{amsmath,amssymb,amsfonts}
\usepackage{algorithmic}
\usepackage{graphicx}
\usepackage{textcomp}
\usepackage{xcolor}
\usepackage{url}
\usepackage{hyperref}
\usepackage{svg}
\usepackage{comment}
\usepackage{pifont}
\usepackage{rotating}
\usepackage{caption} 
\def\BibTeX{{\rm B\kern-.05em{\sc i\kern-.025em b}\kern-.08em
    T\kern-.1667em\lower.7ex\hbox{E}\kern-.125emX}}

\newcommand\chk{\color{black}{\ding{51}}}
\newcommand\crs{\color{red}{\ding{55}}}
\newcommand\crsreason[1]{\color{red}{\ding{55}}\color{black}{#1}}

\begin{document}

\title{Global, robust and comparable digital carbon assets}

\author{
    \IEEEauthorblockN{Sadiq Jaffer\IEEEauthorrefmark{1}\IEEEauthorrefmark{2}, Michael Dales\IEEEauthorrefmark{1}\IEEEauthorrefmark{2}, Patrick Ferris\IEEEauthorrefmark{1}\IEEEauthorrefmark{2}, Thomas Swinfield\IEEEauthorrefmark{1}\IEEEauthorrefmark{3}}
    \IEEEauthorblockN{Derek Sorensen\IEEEauthorrefmark{2}, Robin Message\IEEEauthorrefmark{4}, Srinivasan Keshav\IEEEauthorrefmark{1}\IEEEauthorrefmark{2}, Anil Madhavapeddy\IEEEauthorrefmark{1}\IEEEauthorrefmark{2}}
    \IEEEauthorblockA{\IEEEauthorrefmark{1}Cambridge Centre for Carbon Credits, University of Cambridge
    \\}
    \IEEEauthorblockA{\IEEEauthorrefmark{2}Department of Computer Science and Technology, University of Cambridge
    \\}
    \IEEEauthorblockA{\IEEEauthorrefmark{3}Department of Zoology, University of Cambridge
    \\}
    \IEEEauthorblockA{\IEEEauthorrefmark{4}Lambda Cambridge Ltd
    \\}
}

\maketitle

\begin{abstract}
Carbon credits purchased in the voluntary carbon market allow unavoidable emissions, 
such as from international flights for essential travel, to be offset
by an equivalent climate benefit, such as avoiding emissions from tropical deforestation.
However, many concerns regarding the credibility of these offsetting claims have been raised. 
Moreover, the 
credit market is manual, therefore inefficient and unscalable, and non-fungible, therefore illiquid.
To address these issues,
we propose an efficient digital methodology that combines remote sensing data, modern econometric techniques, and on-chain certification and trading
to create a new digital carbon asset (the PACT stablecoin) against which carbon offsetting claims can be transparently verified.
PACT stablecoins are produced as outputs from a reproducible
computational pipeline for estimating the climate benefits of carbon offset
projects that not only quantifies the CO2 emissions involved, but also allows for similar credits to be pooled 
based on their co-benefits 
such as biodiversity and jurisdictional attributes, increasing liquidity through fungibility within pools. 
We implement and evaluate the PACT carbon stablecoin on the Tezos blockchain, which is
designed to facilitate low-cost transactions while minimizing environmental impact. 
Our implementation includes a contract for a registry for tracking issuance, ownership, and retirement of credits, 
and a custodian contract to bridge on-chain and off-chain transactions. 
Our work brings scale and trust to the voluntary carbon market by providing a transparent, scalable, and efficient framework 
for high integrity carbon credit transactions.
\end{abstract}


\section{Introduction}
To meet the 2015 Paris Agreement's goal of limiting temperature increases to below two degrees
of pre-industrial levels, countries and organisations must rapidly decarbonise and transition towards net zero carbon emissions.
Since there will always be some unavoidable residual emissions, every net zero pathway
incorporates carbon credits that represent
avoided, reduced or removed emissions.
Emitters use these credits to offset
their climate harming emissions. The demand for voluntary carbon market (VCM) credits is
already high and projected to grow from \$2bn in 2023 to \$250bn by 2050~\cite{mckinsey-tsvm-report}. 

Carbon credits aim to balance unavoidable climate harm (e.g. from international air
travel~\cite{gossling2019can}) with an equivalent climate benefit from sequestered~\cite{mcqueen2021review} or avoided~\cite{sohngen2008avoided} 
carbon emissions. 
For this asset class to be credible, however, the claimed climate benefits need to add up to be at least as
effective as the climate harms being offset. 
Unfortunately, the current VCM has numerous credibility gaps due to
the overestimation of climate benefit~\cite{greenfield2021carbon}.
Moreover, it has high transaction and intermediary costs~\cite{hodgson2022surge}, and is subject 
to significant price fluctuations~\cite{twidale_mcfarlane2023carbon}, 
which make the long term investments in credit-generating projects hard to sustain, 
particularly for nature-based projects~\cite{sohngen2008avoided} which can require decades to realise their climate benefit~\cite{karolyi2023biodiversity}.

Our research goal is to use the emerging technology of digital ledgers and permissionless blockchains to solve this problem. 
Specifically, recently, fiat-backed tokens, known as ``fiat stablecoins'', have become widely deployed\cite{baughman2022stable}.  The largest, Tether, has a market capitalisation of nearly \$100bn as of January 2024\cite{sandor2024tether}.
Fiat stablecoins offer cross-border transactions at minimal cost with low volatility.
These stablecoins have three useful digital properties:
\begin{itemize}
    \item \textbf{Self-adjusting valuation}: the value of a fiat currency is determined only by the supply and demand against other currencies and commodities
    \item \textbf{Fungibility}: individual coins can be transparently interchanged for each other 
    \item \textbf{Liquidity}: sufficient quantities of stablecoins are usually available for transactions to clear quickly and at low cost
\end{itemize}
These properties are useful for carbon credits, as well. Self-adjusting valuation allows the value of a project to be determined by the market,
taking into account the current knowledge state of project implementation, rather than set by the project owner by fiat,
thus increasing trust in this valuation.
Fungibility allows credits to be traded across projects, rather than having a project-linked credit that 
cannot be compared with credits from other projects.
Liquidity allows the owner of a carbon credit to feel secure that their investment can be recouped by sale in the market with a low
transaction fee. 

For these reasons, this paper discusses the design and implementation of the PACT stablecoin and its use to provide scale and trust 
in the VCM.
Specifically, we define a structure for creating PACT stablecoins -- digital tokenised carbon assets that are open to scientific examination, auditable on a public ledger, 
and comparable against each other.
This provides a basis for scaling the VCM
to allow participants to offset their necessary emissions efficiently,
which will result in more finance being injected to increase the {\em supply} of nature-based protection interventions, which is urgently needed to stem the devastation of natural environments due to anthropogenic actions~\cite{goudie2018human}. 

The contributions of our work are as follows:
\begin{itemize}
    \item We present the design of the PACT carbon stablecoin 
    \item We discuss the implementation of our stablecoins as on-chain tokens using a reproducible computational pipeline using public trusted data and smart contract-based registries
    \item We demonstrate how the PACT carbon stablecoin addresses the scale and trust issues in the voluntary carbon market
\end{itemize}
The remainder of the paper is laid out as follows. Section \ref{sec:bg} provides the necessary background to our work. In section \ref{sec:architecture} we present the design of the PACT stablecoin which addresses the properties discussed earlier in this section. Sections \ref{sec:proposed-solution} and \ref{sec:implementation} detail and evaluate our implementation of the PACT stablecoin respectively. We then discuss how our work relates to existing initiatives and future opportunities in section \ref{sec:discussion}.

\section{Background}
\label{sec:bg}

\subsection{Problems in the Voluntary Carbon Markets}

The Taskforce on Scaling Voluntary Carbon Markets, the precursor to the Integrity Council for Voluntary Carbon Markets (ICVCM), along with McKinsey estimated the market for voluntary carbon credits could reach \$50bn by 2030~\cite{mckinsey-tsvm-report}. Despite this there are widespread concerns that credits from REDD+ (Reduced Emissions from Deforestation and forest Degradation) projects, which make up more than 40\% of the voluntary carbon market~\cite{unredd2022forest}, overestimate their climate benefit~\cite{greenfield2021carbon,greenfield2023revealed} due to pessimistic baselines. The remaining $\approx$40\% of credits sold in 2022 were Renewable Energy credits~\cite{climatefocus2023vcm}, the additionality of which are unclear with the financial competitiveness of renewables~\cite{lakhani2023revealed}. Verra, the largest certifier of carbon credits on the voluntary market with over 70\% market share \cite{climatefocus2023vcm}, have several avoided deforestation methodologies VM0007~\cite{verra2023reddmf}, VM0015~\cite{verra2023vm0015} and VM0009~\cite{verra2023vm0009} -- all of which afforded project developers some freedom in creating their own baselines. Verra have committed to phasing out these methodologies by 2025~\cite{guardian2023biggestcarbon}.

Baselines chosen on a per-project basis result in wildly different additionality and are thus not \textit{comparable}. This makes REDD+ credits issued under Verra methodologies difficult to value, makes them non-fungible and thus limits available liquidity. This has led to concerns that the voluntary carbon market is becoming a lemons market ``where buyers have no way of distinguishing quality, so some sellers flood the market with bad products, leading to a breakdown of trust and ultimately market collapse.''~\cite{lewsey2023carbon}.

\subsection{Carbon Credit Valuation}
\label{s:value}

The problems with the existing markets lead us to applying econometric techniques to the problem of valuing carbon credits, which we will explain next. Each carbon credit stems from an intervention project that either removes or avoids the emission of CO2. 
An example of an avoided emissions intervention is to replace palm oil plantations which clearcut tropical rainforest with mixed-forest cocoa plantations that have lower yield but preserve the forest environment~\cite{sohngen2008avoided,mbomaalternative}.
The project has a base cost in the form of  the resources required to apply these incentives to shift the local economy away from deforestation. 
The project is then analysed for the additional tonnes of avoided CO2e (Carbon Dioxide Equivalent) from the intervention for a resulting \$/CO2e tonne. 
Calculating the additional tonnes involves quantitatively and reliably determining the following values, each of which requires specialised inspection of the nature of the intervention involved.

Creating carbon stablecoins presents different challenges from conventional fiat stablecoins, which we examine next.

\subsubsection{Additionality}

The additionality of a project is the CO2e removed or avoided by the project's intervention~\cite{schneider2009assessing}. This is
determined against a business-as-usual case or a baseline via a counterfactual analysis~\cite{bennett1987event,swinfield_balmford_2023}. For some project
types, such as Direct Air Capture~\cite{mcqueen2021review}, determining this baseline is relatively simple -- one is unlikely to
pull CO2 from the atmosphere and sequester it without being paid to do so. For others though, especially
nature-based projects such as avoided deforestation, the choice of baseline is more complicated. Overly
pessimistic baselines can lead to over-estimates of project CO2e benefits~\cite{guizarcoutino2022,west2023},
whilst underestimation make many projects financially unviable. To further complicate valuation, projects from
the same jurisdiction may use differing methods for determining their baselines~\cite{teo2023uncertainties}.

\subsubsection{Leakage}

Project interventions may cause unintended climate damage outside of their project area. Consider an avoided
deforestation project that simply shifts deforestation to the area surrounding the project -- this has ultimately
not led to an avoidance of carbon being emitted into the (shared!) atmosphere~\cite{wunder2008we}. Leakage can be {\em local}, as in the previous example, or it can be {\em global} where forgone production from the intervention leads to sustained demand shifting supply of that commodity to elsewhere in the world~\cite{gan2007measuring}. Accounting for local leakage can be assessed using data from the areas surrounding the project. On the other hand, global leakage is a much harder problem as it requires knowledge of the specific interventions carried out by the project to account for e.g forgone production of a particular commodity. 

\subsubsection{Permanence}

Many classes of projects are \textit{impermanent}; the net CO2e removed or avoided has a risk of being later released back in to the atmosphere either during or after the end of the project~\cite{herzog2003issue}. This is especially true of nature-based projects where natural ecosystems are always vulnerable to net loss disturbances such as fire, disease or floods. We therefore need compare between projects where one offers permanent carbon sequestration but the other offers a potentially larger but impermanent emissions avoidance.  When a portfolio of projects with different levels of impermanance are pooled, we need a mechanism to make these comparable or else they represent different levels of climate benefit.

\subsection{Carbon Credit Fungibility}
\label{sec:fungibility}

Many intervention projects represent augmented or alternative livelihoods to local people, and thus offer {\em co-benefits} beyond just their value in removing or avoiding CO2e. These co-benefits range from increasing or avoiding the reduction of biodiversity and natural habitats~\cite{phelps2012biodiversity}, to increasing the prosperity of local inhabitants~\cite{fischer2023community} and the promotion of equality and fair redistribution of wealth in the region~\cite{godden2016redd}. For most buyers of carbon credits, projects offering these co-benefits can attract significant premiums since they help to balance a portfolio of worldwide interventions against a variety of environmental, social and corporate governance (ESG) factors.

\begin{table*}
\centering
\begin{tabular}{|l|c|c|c|c|c|c|c|}
\hline
\textbf{Project} & \textbf{AL$_{adj}$} & \textbf{eP} & \textbf{£$_{project}$} & \textbf{£$_{PACT}$} & \textbf{Biodiversity} & \textbf{Livelihood} & \textbf{Justice} \\ \hline
Climeworks & 1 & 1 & 900 & 900 & C & B & A \\
CarbonCure & 1 & 1 & 145 & 145 & C & B & A \\
African rainforest conservation & 0.23 & 0.35 & 5.8 & 73 & A & A & A \\
Ecuadorian rainforest conservation & 0 & 0.20 & 15 & --- & A & B & B \\
Mexican tree planting & 0.48 & 0.36 & 17.5 & 101 & B & A & A \\
University of Cambridge Woodland Creation & 1 & 0.77 & 112 & 145 & B & B & A \\
UK Fenland conservation & 1 & 0.75 & ? & ? & C & B & B \\ \hline
\end{tabular}
\caption{PACT evaluation of several project interventions (reproduced from~\cite{swinfield_balmford_2023})}
\label{tab:pact_table}
\end{table*}

A carbon asset should ideally be interchangeable across different interventions, since it ultimately represents some tonnes of CO2e that are avoided or removed, as calculated from the additionality, leakage and permanence values of the project (\S\ref{s:value}). However, the co-benefits of the project illustrate that in reality projects differ through the interventions used on the ground, their effectiveness and the jurisdiction(s) in which they take place. 
This matters hugely in practice since any credit-generating project is subject to a risk of reversal or over-crediting, and so a single controversial project in a larger pool of carbon credits can inflict reputational damage to a well-intentioned purchaser~\cite{song2019even}. Since the carbon markets are currently voluntary, this blowback leads to adverse selection effects in the market, and the preferred action for well-intentioned purchasers is therefore often inaction (i.e. non-participation in carbon credits) that hinders wider adoption of offsetting.

It is therefore essential that carbon stablecoins are not only accurate representations of the quantitative emissions benefits, but that they also track the qualitative co-benefits and allow for buyers to define their own thresholds for fungibility. One buyer may value biodiversity preservation much more than another buyer, and therefore not treat otherwise-equivalent (from an emissions perspective) assets equivalently.

\subsection{Carbon Credit Liquidity}
\label{sec:liquidity}

It is not unusual for a carbon credit-generating project to last for decades, especially those based around
natural climate solutions. Project developers thus often face delays of years between starting a project and realising the climate benefit which allows carbon credits to be issued and sold~\cite{peskett2011institutional}.
If trees are planted, or forests are not cut down, it can take 5-10 years from the inception of the project to verify the initial benefit of avoided emissions from net forest carbon stock. Therefore, the existing carbon market is extremely supply constrained due to the unusual upfront financing needs~\cite{van2018contribution}.
Another differentiator from fiat stablecoins, is that the available pool of carbon stablecoins steadily decreases due to the fact that every carbon stablecoin can only be spent once (dubbed as ``retired'') to offset climate damage, which ensures that there is a further constant pressure on supply.

Meanwhile, the demand for carbon credits is forecast to grow to \$250bn by 2050\cite{mckinsey-tsvm-report}. Left to itself, this will cause the price to soar (leading to organisations and individuals not being able to offset) or for junk supply of credits  (resulting in, at best, minimal climate benefit). The other (more desirable) direction is that a carbon credits market dramatically increases the {\em supply} of projects available by incentivising early financing for promising projects, but without making future promises about emissions that may not pan out in practise.

\section{Architecting our PACT carbon stablecoin}
\label{sec:architecture}

We have so far examined the differentiating properties -- both quantitative and qualitative -- needed from a carbon stablecoin. We next examine how we overcome these complexities and enable carbon credits to be valued effectively and transacted at low cost whilst still acknowledging their respective co-benefits. 
We dub our carbon stablecoin a PACT (for ``Permanent Additional Carbon Tonne'') as it systematically accounts for the permanence, additionality and leakage (\S\ref{s:value}) for every project~\cite{Balmford_Keshav_Venmans_Coomes_Groom_Madhavapeddy_Swinfield_2023}.

\subsection{PACT Valuation}
\label{sec:pvaluation}

Currently, carbon projects self-declare their own background baselines, and then calculate their additionality against that baseline. This is problematic since different projects would have declared different baselines, and thus would not be comparable (and their valuations in fiat would be accordingly skewed). This leads to a huge range of prices per tonne in the current market. Our carbon stablecoin design instead uses global, comparable baselines by deploying modern econometric techniques~\cite{swinfield_balmford_2023}  that can be measured digitally for a class of project intervention (e.g. satellites for avoided deforestation or restoration forestry projects). This then implies that any set of carbon credits can be quantitatively compared regardless of their intervention type.

The PACT comparability property allows for robust relative valuation between projects; if a single permanent additional carbon tonne (PACT) of Direct Air Capture is \$1000 then a REDD+ avoided deforestation credit~\cite{rakatama2017costs} offering the same at \$150 indicates one is over- or under-priced. An example set of projects can be found in Table~\ref{tab:pact_table}. The $AL$ and $eP$ represent additionality, leakage and permanence results, and $\pounds_{project}$ is the base price per tonne of the project. $\pounds_{PACT}$ represents the adjusted and comparable price across projects. The remaining columns show the ratings for the project -- textual analysis is attached to the PACT metadata. We go into more detail about these calculations later (\S\ref{sec:implementation}).

We also use digital measurement reporting and verification (dMRV) where possible (e.g. satellites for forests) to give us a global basis on which to calculate the valuation. A reproducible calculation infrastructure then acts as a transparent oracle, enabling data feeds into a blockchain- or ledger-based tokenisation platform. Without all of these pieces, carbon credit values are prone to significant volatility and risk of being junk credits due to methodological or measurement errors.
Together, however, these valuation techniques allow carbon projects to be significantly more reliably valued in terms of price per tonne.

\subsection{PACT Fungibility}
\label{sec:pfungibility}

Quantifying projects against comparable baselines transitively enables their eventual price (which combines the base cost of the intervention) to also be comparable.
However projects -- especially those that are nature-based -- come with significant co-benefits such
as improvements to biodiversity and local livelihoods which need to be qualitatively analysed and scored
more broadly than the precision required of CO2e calculations. These co-benefits must be tracked with each
PACT issued by the project, so that the {\em buyer} can filter pooled PACTs on the basis of their threshold
for these co-benefits, reflecting their own risk appetite and strategic ESG drivers.

PACT carbon stablecoins are therefore augmented with additional metadata that:

\begin{itemize}
\item categorises them based on qualitative ratings, much like bonds. These ratings are
simple A/B/C rating and come with associated textual descriptions. Common categories
include biodiversity preservation ratings, local livelihood impact, and justice frameworks that
check topics such as free-and-informed-consent for indigenous inhabitants.
\item defines the political jurisdiction in which a project takes place. Many buyers wish to
spread out their portfolio of carbon credits across geographies to minimise risk of reversal
(e.g. due to wildfires)~\cite{chingono2023zimbabwe}.
\item describes the source of funding for the carbon project, to facilitate ``greenwashing'' checks
for buyers that do not wish to be associated with certain sources of funds (e.g. from fossil-fuel processing).
\end{itemize}

A pooled token can credibly reflect a PACT (permanent additional carbon tonne) \textit{across}
different projects, and buyers can also select their contents based on criteria that matter to them (such as a
minimum biodiversity benefit).
Thus with PACTs pooled by their co-benefits and jurisdiction, we can achieve fungibility within pools.

\subsection{PACT Liquidity}
\label{sec:pliquidity}

With PACTs pooled according to co-benefits and jurisdiction, risk can be spread across the pool
much more robustly.
Once the overall market risks drops, it becomes more viable to justify the financing of new supply
of carbon credits via new interventions, which would result in significant numbers of tokens backed by each pool being available.
This fungibility can then enable liquidity on decentralised financial infrastructure such as automated market makers (AMMs).

Increased liquidity on digital trading infrastructure opens up new uses, such as allowing real-time carbon offsetting and on-demand offsetting with immediate retirement.
This is currently not possible on conventional carbon markets, since the purchase of credits is a time-consuming process that involves bilateral negotiations with brokers and careful per-project checking to minimise the risk of reputational damage.

\section{Implementing the PACT carbon stablecoin}
\label{sec:proposed-solution}

\begin{figure}
    \centering
    \includegraphics[width=1\linewidth]{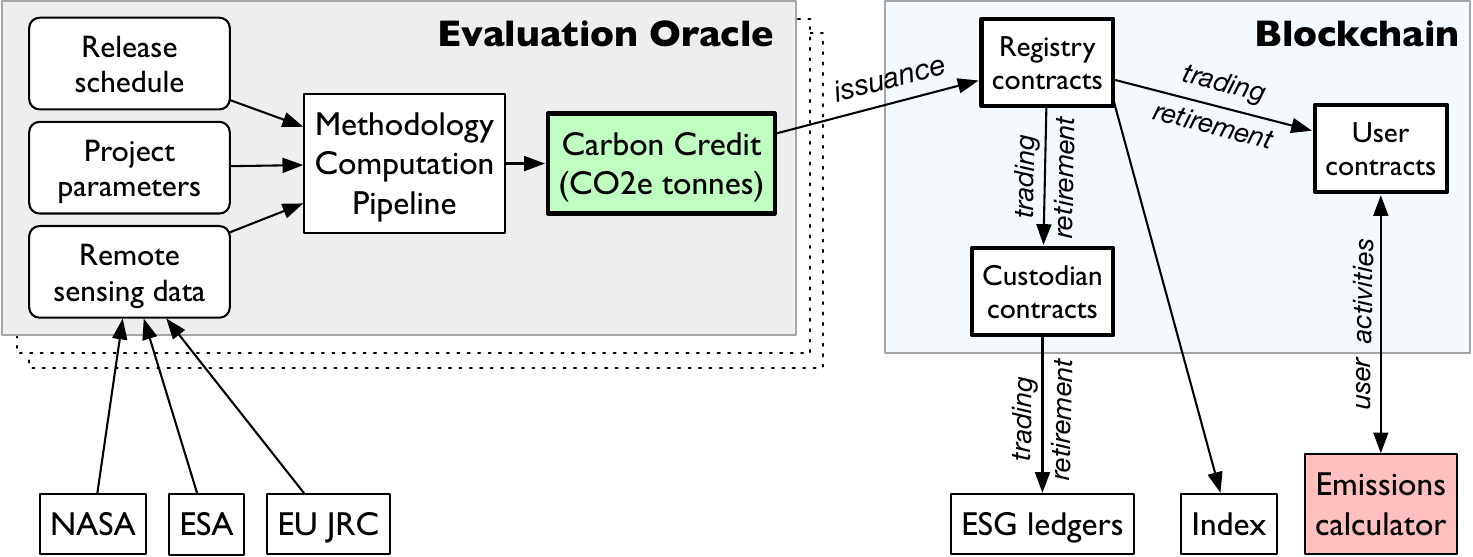}
    \caption{System architecture showing the evaluation oracle feeding carbon credits (climate benefit) into a smart contract system. Clients can depend on oracles to calculate emissions harm, and indexers can track global progress.}
    \label{fig:architecture-diagram}
\end{figure}
To address the valuation, fungibility and liquidity of carbon credits, we construct a solution comprised of several components:
\begin{itemize}
    \item Reproducible computational pipelines that mechanise the evaluation of projects enabling transparent checking of calculations by third parties
    \item Pooling mechanisms that group underlying projects by their co-benefits permitting fungible tokens that increase liquidity and enable automated trading
    \item Emissions-efficient carbon credit issuance and retirement tracking via a smart contract-based registry that permits for both on- and off-chain trading
\end{itemize}

We now describe how we implement the PACT stablecoin architecture (\S\ref{sec:architecture}) as on-chain tokens in our experimental prototype (see Figure~\ref{fig:architecture-diagram}).
Before considering the individual components of the proposed system, we first introduce the econometric techniques and frameworks we have built to make this possible.

\subsection{Algorithmic quantification of project climate benefit}
\label{sec:algorithms}

The short form to calculate a PACT value is $(A-L) \times eP$, where the project additionality $A$ is adjusted by leakage $L$ and multiplied by the equivalent permanence $eP$. That formula is a general ex-post framework for quantifying the climate benefit of a given project within a specified timeframe, described in detail in the literature~\cite{swinfield_balmford_2023,Balmford_Keshav_Venmans_Coomes_Groom_Madhavapeddy_Swinfield_2023,balmford2023pact}.
To summarise, the method involves a two-step calculation process.

\subsubsection{Calculation of Additionality and Leakage}
A project is defined as a set of polygon boundaries within a given jurisdiction. The first step involves determining the additionality and leakage of the project area, i.e., the additional climate benefit generated as a direct result of the project and unintended climate damage observed as a consequence of the project. This is achieved by comparing the observed outcomes with a statistical counterfactual, which represents the business as usual case.

\subsubsection{Adjustment to equivalent Permanence}
The next step adjusts the net additionality so that an impermanent intervention can be viewed as to be equivalent to geologically permanent carbon sequestration. This is achieved through the use of an ex-ante but conservative release schedule which models the rate at which the net additional carbon in the project is released in to the atmosphere.

As the additionality calculation is ex-post, the statistical counterfactual is constructed with observational ground-truth data that permits for a mechanised, reproducible digital process.

\subsection{Integration of blockchain data oracles}

The mechanisation of the $(A-L)*eP$ evaluation allows for the implementation of a fully digital evaluation pipeline that functions as a data oracle~\cite{chainlink2023blockchain}. The evaluation pipeline processes and publishes evaluation data on a per-project basis in such a way that all of the original inputs are tracked, ensuring transparency and verifiability throughout the project evaluation process.

For each project, the pipeline publishes and pins the following components to the InterPlanetary File System (IPFS)~\cite{benet2014ipfs}.

\subsubsection{Ground Truth Data}
All data used for the construction of statistical counterfactuals used in additionality and leakage calculations. This puts restrictions on the types and availability of the data used as this necessarily needs to be freely available under a license that permits for reproduction.

\subsubsection{Statistical Counterfactuals}
The calculated counterfactuals are made available. Whilst they could be reconstructed from the ground truth data, the step of producing them is computationally intensive and so their availability permits for lighterweight verification, as well as additional tooling that allows for visualisation.

\subsubsection{eP Release Schedules}
As each methodology and potentially project has their own release schedule used for the calculation of equivalent permanence, these must be made available for reproduction as well as annual reccalculation of permanence and credit adjustment.

The hashes of these artefacts are recorded alongside the evaluation results, ensuring immutability and traceability of a given evaluation resulting and also permitting independent third-party verification by reproducing the (open source) computation pipeline.

\subsection{Mechanisms for pooling co-benefits}

As described earlier (\S\ref{sec:pfungibility}), in addition to their primary climate benefit, projects often yield significant other co-benefits including improvements in biodiversity, local livelihoods, and social justice.
Our pipeline tracks co-benefits as follows.

\subsubsection{Quantification of co-benefits}
Where there are robust methods to do so, we quantitatively assess the co-benefits of projects, which allows for comparison between projects. If a numerical assessment is not yet feasible, then a qualitative assessment is sufficient and assigned an A/B/C rating by human experts. Our most commonly tracked co-benefits include biodiversity, livelihoods and justice -- while there are emerging mechanisms for quantifying biodiversity~\cite{eyres_life_2023} and livelihood impact~\cite{burke2021using}, the most robust state-of-the-art for these is still qualitative assessment.
\subsubsection{Categorisation and Pooling}
Projects are categorised into groups based on their expert-rated co-benefits. This permits the construction of distinct PACT tokens each backed by pools containing credits from evaluated projects with similar co-benefits. These tokens, denominated in CO2e tonnes, are thus fungible whilst preserving the value of project co-benefits.

Through this approach we can balance the benefits of fungibility whilst still reflecting the value of project co-benefits.

\subsection{Credit issuance and tracking}

To manage issuance, trading, and retirement of PACT carbon credits in an emissions-efficient way, we use a smart contract-based blockchain system. We are pragmatic as to the manner in which the majority of credits are transacted currently and so the system is designed to support both on-chain and off-chain trading activities. The PACT system comprises three key contracts.

\subsubsection{Central Registry Contract}
At the core is a central registry contract responsible for tracking the issuance, retirement and ownership of credits. Ownership is tracked in this contract against an on-chain wallet or smart contract address, facilitating integration with existing blockchain token interfaces (e.g FA2 on Tezos or ERC20 on Ethereum) and thereby enables interoperability with wallets and decentralised finance (DeFi) systems.

The registry also includes a ``retirement'' function to enable the offsetting of credits against climate damage. With the potential requirement for billions of tonnes of carbon credits~\cite{mckinsey-tsvm-report} by 2030, we need an efficient way retire credits that does not itself cause significant carbon emissions. While any retirements system could be made more emissions efficient by coarsening the units of retirement, this hinders detailed checking of organisation's offsetting claims. Thus, our mechanism for retirement minimises the additional on-chain storage required due to the significant carbon footprint involved in on-chain storage\cite{tannu2023dirty} by providing a specialised retirement entrypoint (and not, for example, issuing NFTs per retirement, which would increase the on-chain storage requirements).

\begin{figure}[h]
    \centering
    \includegraphics[width=0.85\linewidth]{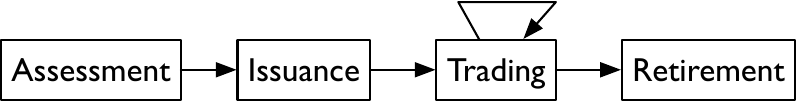}
    \caption{Stages in carbon credit lifecycle}
    \label{fig:credit-lifecycle}
\end{figure}

\subsubsection{Pooled Credit Contract}
A specialised contract is proposed for pooling project credits by their categorised co-benefits, this creating pooled PACT credits. This contract mirrors the interface of the central registry, ensuring interoperation with existing blockchain and DeFi infrastructure and enabling the management and trading of pooled credits.
\subsubsection{Custodian Contract}
To bridge the gap between on-chain and off-chain financial systems, a 'custodian' contract is proposed. This contract is designed to hold both individual project PACT credits and pooled PACT credits on behalf of off-chain entities. It serves as a link, allowing traditional finance systems to engage in trading activities that are reflected and tracked on the blockchain.

Through this system we aim to establish an emissions-efficient and transparent system for the management of carbon credits. This approach not only addresses the immediate needs of credit issuance and retirement but also pragmatically integrates with existing financial ecosystems, thus making wider-adoption a more realistic prospect.

\section{Evaluating the PACT carbon stablecoin}
\label{sec:implementation}

\begin{figure}
    \centering
    \includegraphics[width=0.99\linewidth]{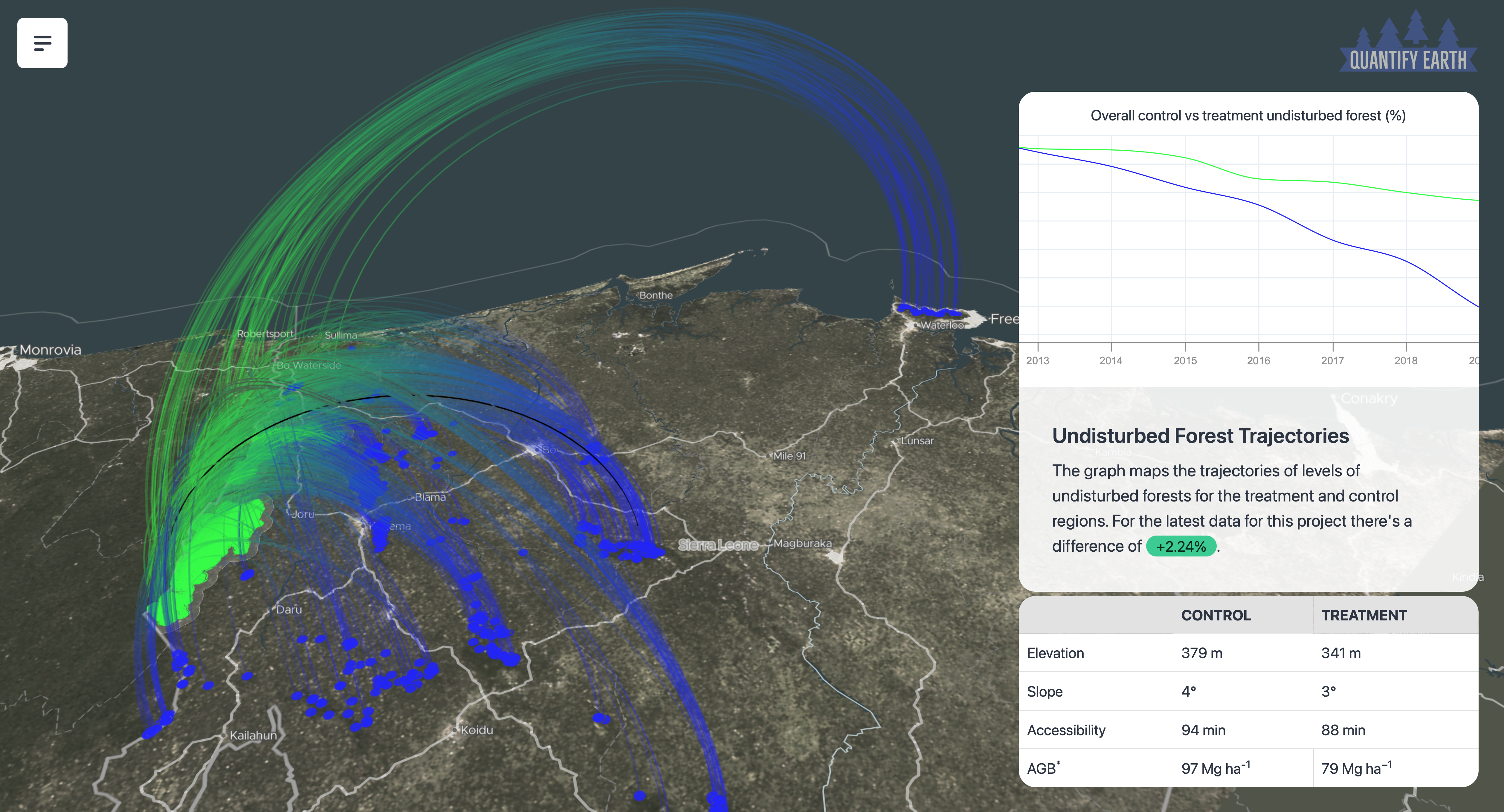}
    \caption{A visualisation of the counterfactual pixel-matching pipeline to assess the additionality of an avoided deforestation project in Sierra Leone. Pixels in green are from the project and they are matched to similar locations in blue.}
    \label{fig:qe}
\end{figure}

In order to test the efficacy of our architecture, we have built a \href{https://github.com/quantifyearth/tmf-implementation}{prototype} of the PACT system architecture (\S\ref{sec:proposed-solution}), with a focus on avoided deforestation projects for tropic rainforests. Our implementation work has been focused on three core areas, each corresponding to a key component of the proposed solution.

\subsubsection{Development of an Evaluation Pipeline}

We have developed an \href{https://github.com/quantifyearth/tmf-implementation}{open-source software pipeline} that evaluates suitable projects using this approach. The pipeline serves as the core of the proposed data oracle, capable of processing and publishing project data. It is designed to allow for full reproducibility, ensuring that third parties can independently verify and replicate evaluations. Figure~\ref{fig:qe} shows an example screenshot from the evaluation project browser, which is one of the forestry projects from Table~\ref{tab:pact_table}.

We implemented the PACT Tropical Moist Forest methodology~\cite{balmford2023pact} for our pipeline.
The methodology uses multiple sources of publicly available data to make the assessment based on comparison tens of thousands of treatment pixels (samples) in both the project zone and its surrounding leakage zone to corresponding paired control pixels in the surrounding region. Pairing is done based on similarity of physical properties of the pixels (country, biome type, elevation, slope, and proximity to human habitation), along with similar historic behaviour in terms of land usage~\cite{vancutsem2021} up until the project start point. Once the project starts, we then examine how the land usage changes over time for the treatment pixels versus the counterfactual control pixels, which indicate behaviour had the intervention not taken place.  By combining the local land usage data with AGB data~\cite{DUNCANSON2022112845}, we calculate the carbon density of the area and assess both the additionality of the project area (i.e., carbon saved compared to the control counterfactual pixels) and leakage (signs of activity displaced into a buffer around the project) to create an overall permanence evaluation for the project.

To enable adoption of the assessment methodology, we have provided a reference pipeline implementation written in Python and controlled by the OCurrent workflow library. The bulk of the work is done a series of Python scripts, each performing one step in the methodology: the language choice and the breaking down of the methodology into individual steps was made to make the implementation scrutable to ecologists. These individual steps are then orchestrated by an OCaml-based workflow engine, which provides a way to ensure the scripts not only run in the right order, but unnecessary repeated work is avoided, by tracking result dependancies we can avoid re-computing old results where those individual scripts or their inputs don't update. A typical computation pipeline where a large project is involved can take up to a day to process on a fast machine.

\subsubsection{Smart Contracts}
The blockchain piece of our implementation is a suite of \href{https://github.com/carboncredits/x4c}{open-source contracts on the Proof-of-Stake Tezos blockchain}~\cite{goodman2014tezos}. These contracts implement the registry and custodian functionalities, designed to manage the issuance, trading and retirement of carbon credits efficiently and securely. The choice of the Tezos blockchain and implementation reflect the requirements of minimising the environmental impact of the proposed systems.

We chose the Tezos blockchain for the implementation of our proposed system due to its low carbon footprint\cite{tezoslca2021} and potential to deliver significant transactional throughput for carbon stablecoin transactions at low gas cost~\cite{nomadic-scaling}.
Tezos also has a novel self-amending distributed governance mechanism built into the chain, allowing for technological upgrades to be voted on by users~\cite{allombert2019introduction}. This mechanism builds confidence in the longevity of the underlying blockchain itself to run for decades, to match the carbon credits projects that it is tracking.

The \href{https://github.com/carboncredits/x4c/blob/main/src/fa2.mligo}{registry contract} performs the following key functions:
\begin{itemize}
    \item \textbf{Issuance:} The registry issues PACT tokens on instructions from the data oracle associated with the particular contract instantiation
    \item \textbf{Tracking of ownership:} Transfer of PACT tokens between on-chain entities and the querying of balances
    \item \textbf{Retirement:} Tokens can be \textit{burned} in order to retire and offset them. Each retirement can only happen once.
\end{itemize}

The registry also implements the Tezos FA2 token standard~\cite{gabbay2021money} which enables interoperability with existing Tezos wallets, marketplaces and automated market-makers (AMMs). Each instance of the contract on-chain is associated with a single data oracle that can mint and update contract metadata. The contract allows for transfer of oracle to provide for evaluation updates.

Each contract contains one or more token types, numerated by their \textit{token id}. The contracts contain a metadata mapping against each token id, which in the proposed system would contain immutable references to data on the project itself, geographic information, project durations and finally the categorised assessments of co-benefits.

\subsubsection{Issuance Smart Contracts}
\label{sec:issuance}

The \textit{mint} entrypoint enables the oracle to issue PACT tokens to an on-chain address which are added to the addresses' balance in the contract. A transactional call to the mint entrypoint, e.g from a project developer issuing a new set of credits, can contain structured metadata and this is the mechanism used for recording the hashes of the evaluation source code, its inputs and outputs. This data is ``emitted' by the smart contract but not added to on-chain storage -- this means it will be accessible from chain explorers and archival nodes. This mechanism, in contrast to on-chain storage, does not necessitate fast rapidly-accessible storage and so significantly lowers the emissions involved in minting. This same mechanism is used for retirement.

The contract contains two entrypoints for tracking of token ownership: \textit{Transfer} and \textit{Balance}. Transfer moves tokens between the balances of two on-chain addresses and balance simply returns the total balance of a given token at an address, in the contract. Due to adhering to the FA2 token specification, the contract also supports \textit{operators} which enables delegating authority to transfer tokens belonging to an address to another address. This is similar to the the allowance functionality in Etheruem's ERC20\cite{erc20eth}.

\subsubsection{Retirement Smart Contracts}

The retirement entrypoint allows an address (or their operator) holding PACT tokens in the registry to destroy part of their balance and do so with an associated set of metadata. This is used to ``offset'' the tokens against a claimed carbon emission and through specific details in the metadata as to the cause of that carbon emission organisations can more precisely show their use of offsets is in addition to their decarbonisation work.

Retirements of PACT tokens are designed to be \textit{emissions efficient}, that is they minimise the climate damage involved in the retirement itself. To do this they ``emit'' metadata during retirement operations, which makes the data available to indexers and archival nodes but does not occupy on-chain storage.

\subsubsection{Custodian Smart Contracts}

The \href{https://github.com/carboncredits/x4c/blob/main/src/custodian.mligo}{custodian contract} is intended to act as a bridge between on- and off-chain transactions of PACT tokens. Each instance of the custodian contract enables an on-chain entity to hold tokens on behalf of off-chain entities and issue, manage and retire on their behalf. An off-chain entity is represented by a set of Know Your Customer (KYC) metadata specific to the custodian instance. It has the following main entrypoints:

\begin{itemize}
    \item \textbf{Transfer internal/external:} Transfers PACT token balances between two off-chain entities (internal) or from the custodian to an on-chain entity (external). Trades of PACT tokens between off-chain entities (e.g through an existing financial exchange or bilaterally) are required to be reported to the custodian, who in turn issues an internal transfer of the tokens between the entities in the smart contract.
    \item \textbf{Mint/Retire} Facilitates issuance and retirement of PACT tokens to an off-chain entity, such as a project developer or corporate buyer. These are ultimately carried out at the registry, with the custodian acting as a proxy.
\end{itemize}

Thus the custodian contract enables interoperation with existing carbon credit trading venues, facilitating the reporting of transactions on-chain and maintaining transparency.

\section{Discussion}
\label{sec:discussion}

The combined properties of the proposed PACT carbon stablecoin enable it to address the scale and trust issues currently plaguing the voluntary carbon market.

Scale is achieved through:
\begin{itemize}
    \item data-driven project baselines that use freely available remote sensing data, enabling low-cost project evaluations and monitoring
    \item automated evaluation pipelines that serve as data oracles for the issuing of on-chain tokens
    \item pool mechanisms that group underlying projects and permit fungible tokens that increase liquidity and enable automated trading
    \item smart contracts that provide on-chain emissions-efficient registration and custodianship of credits
\end{itemize}

We believe trust is also achieved via:
\begin{itemize}
    \item statistical and robust project baselines derived from remote sensing data and thus avoids pessimistic baselines provided by projects themselves
    \item reproducible open-source computational pipelines for evaluation and monitoring that permit cross-checking by third parties
\end{itemize}

\begin{table*}[h]
\centering
\begin{tabular}{l|l|l|l|l}
 Name & \multicolumn{1}{l|}{Comparable for valuation} & \multicolumn{1}{l|}{Fungibility} & \multicolumn{1}{l}{Liquidity}  \\ \hline
 Verra &         \crsreason{(incomparable baselines)}                                 &         \crs            &  \crs \\ \hline
 Toucan &        \crsreason{(incomparable baselines)}              &           \crsreason{(pooled but incomparable)}           &   \chk \\ \hline
 Flowcarbon &        \crsreason{(incomparable baselines)}                &           \crsreason{(pooled but incomparable)}           & \chk \\ \hline
 Moss &        \crsreason{(incomparable baselines)}              &           \crsreason{(pooled but no co-benefits)}           &   \crs \\ \hline
\end{tabular}
\caption{Related carbon asset systems}
\label{tab:related-work-table}
\end{table*}

\subsection{Related Work}
\label{sec:background}

Table \ref{tab:related-work-table} shows several of the most prominent initiatives to bring carbon credits on to various blockchains. The largest is Toucan protocol\cite{ToucanWhitepaper} which tokenizes existing carbon credits in off-chain registries via a bridge. It originally provided this for Verra credits though this was prohibited in 2022\cite{mulder2023asverra}. It currently bridges credits from Puro Earth\cite{puroearth2023}, a newer voluntary standards body that only provides carbon removal credits that have at least 100 year permanence. This precludes nature-based projects such as Avoided Deforestation or Afforestation, Reforestation and Revegetation (ARR) which form major parts of many countries climate-change goals, especially in the global south.

For bridged Verra credits, Toucan pools tokens in two separate pools: Base Carbon Tonne (BCT) and Nature Carbon Tonne (NCT). While this does create fungible BCT and NCT tokens, thus improving liquidity, it has significant drawbacks. The credits forming both pools are not \textit{comparable} - they have differing baselines and permanence \cite{SorensenTokenizedCarbonCredits}. This makes valuing the credits a complex, manual process and indeed, has resulted in both pools containing significant numbers of non-additional "junk" credits \cite{badgleycullenward2022zombies}. Several other projects, such as Flowcarbon and Moss also provided credits from Verra before being prohibited from doing so.
Table~\ref{tab:related-work-table} shows how these tokens compare to the PACT stablecoins, particularly around their comparability.

\subsection{Future Work}
There are many potential areas for future work on the proposed system, in addition to the significant amounts of implementation work still to be carried out.

\subsubsection*{Retirement}

Carbon offsets should be used only for climate damaging activities that are unavoidable and which can not currently be decarbonised \textit{or} historic emissions. That currently this is not always the case opens up even good actors to accusations of "greenwashing". In order to avoid these reputational risks many organisations simply choose not to offset but this need not be the case if there were systems available to enable these organisations to transparently show they were using offsets correctly. It may be possible to use techniques like commitment schemes and verifiable computation to attach metadata to retirement of offsets that enables scrutiny of their use while revealing little information that could be of use to competitors.

\subsubsection*{Further decentralisation}

As proposed the current system centralises the evaluation of projects, the assessment of co-benefits and the issuance of credits with a single data oracle. An alternative design might be that an arbitrary number of unit-less credits are issued for a single project intervention and time period, with evaluations of carbon and co-benefits being added as ``layers'' from multiple participants in the ecosystem. Under this model, one's available carbon offsets in PACTs would be the PACT evaluation layer for the project's intervention in that time period multiplied by one's fraction of total issued tokens for that project intervention and time period. With multiple evaluation layers on a project, buyers could have significant confidence in the true carbon reduction or removals from the project, as well as the co-benefits. There are many open questions for how this market could be structured and the various incentive mechanisms that would need to be in place to encourage evaluators to come forward without burdening project developers with greater costs.

\subsubsection*{Improved fungibility}

Whilst categorising projects by their co-benefits and jurisdiction controls for the major sources of value and risk, it could fail to distinguish very high quality projects as there is a necessary trade-off between the broadness of categorisation and the liquidity of the resulting tokens for those pools. Recent research on structured pools\cite{sorensen2023structured} could be one approach but more radical could be to separate co-benefits from the carbon benefits of projects themselves, in a manner similar to Renewable Energy Certificates\cite{epa2023recs}. If an organisation puts a premium on biodiversity but has no current offsetting needs, they could purchase the biodiversity components of existing projects - which ultimately results in a higher return for the project. This may be impractical without mechanisms to quantitatively assess co-benefits however.

\section{Conclusion}\label{sec:conclusion}

We tackle the problem of carbon credit valuation through the adoption and implementation of econometric techniques~\cite{swinfield_balmford_2023,balmford2023pact} for assessing
additionality and leakage through statistical counterfactual analyses that provide dynamic project baselines. We adopt conservative discounting techniques~\cite{Balmford_Keshav_Venmans_Coomes_Groom_Madhavapeddy_Swinfield_2023} for determining the equivalent permanence of emissions reduced or removed by a project. Together these techniques enable projects to be assessed in Permanent Additional Carbon Tonnes (PACTs) which along with the base price per tonne being offered enables \textit{comparable} valuation across very diverse carbon credit projects. Further we developed a reproducible computational pipeline for estimating the climate benefits of carbon offset projects using these data-driven dynamic baselines. This pipeline can serve as a data oracle, enabling reliable and transparent valuation of projects on the blockchain.

With PACT tokens offering comparable climate benefit, we identify the co-benefits attracting buyer premiums (biodiversity, livelihood impact and social justice) as well as the main risk factor of jurisdiction, and pool tokens by their biodiversity/justice/jurisdiction attributes. This within-pool fungibility allows for the issuance of tokens backed by each pool and which can be traded on any exchange or automated market maker (AMM).

In summary, we believe the contributions made by this paper solve key issues that limit the effective use of carbon credits as digital assets, such as on the blockchain. By widening market participation we can ensure that more entities fully balance their climate damage with climate benefit, and together we can limit the worst effects of climate change.

\paragraph*{Acknowledgements} This research was funded by a donation to the University of Cambridge from the Tezos Foundation (NRAG/719). We thank Andrew Balmford, David Coomes, Harriet Hunnable, Eleanor Toye Scott and Bas Spitters.

\bibliographystyle{IEEEtran}
\bibliography{references}

\end{document}